\documentclass[11pt,a4paper,reqno]{amsart}
\usepackage{cite}
\usepackage{multirow}
\usepackage[centertags]{amsmath}
\usepackage{amsfonts}
\usepackage{amssymb}
\usepackage{amsthm}
\usepackage{newlfont}
\usepackage[pdftex]{graphicx,color}
\usepackage{amscd}
\usepackage{verbatim}
\allowdisplaybreaks
\usepackage{multirow}

\theoremstyle{definition}

\newtheorem{example}{Example}

\def\Z{\mathbb Z}

\def\a{\alpha}

\newcommand{\om}{\omega}
\renewcommand{\rho}{\varrho}
\renewcommand{\phi}{\varphi}

\def\be{\begin{equation}}
\def\ee{\end{equation}}

\begin{document}

\title{Group theoretical methods to construct the graphene}
\author[Z. Grabowiecka \&  J. Patera \& M. Szajewska]{Zofia Grabowiecka$^{2}$ 
Ji\v{r}\'{\i} Patera$^{1,2}$ 
Marzena Szajewska$^{3}$}


\begin{abstract}
In this paper, the tiling of the Euclidean plane with regular hexagons whose vertices are occupied by carbon atoms is called the graphene. We describe six different ways to generate the graphene by the means of group theory. There are two ways starting from the triangular lattice of Lie algebra $A_2$ and $G_2$, and one way for each of the Lie algebras $B_3$, $C_3$ and $A_3$, by projecting the weight system of their lowest representation to the hexagons of $A_2$. Colouring of the graphene is presented. Changing from one colouring to another is called phase transition. 
Multistep refinements of the graphene are described.
\end{abstract}

\maketitle


\section{Introduction}\

The discovery of graphene sheets in nature has been in the center of interest for many years. After it was successfully isolated from graphite \cite{discovery}, empirical investigation of graphene has been continued by both scientists and industry partners as it has remarkable properties and significant applications in medicine, electronics, composite materials, energy and many more \cite{prop1,prop2,prop3,prop4}. Graphene is found to be extremely strong while thin, transparent and flexible material that is electrically and thermally conductive. 

Graphene is one-atom thick layer of carbon arranged as vertices of regular hexagons. In this paper we focus on the origin of its structure from the mathematical point of view, more precisely obtaining the hexagonal tiling of a plane  from different symmetry groups.

The simplest prescription for forming a graphene sheet is to take a regular hexagon, reflect it an unlimited number of times in its sides and join obtained hexagons by their sides. This is a simple construction, which offers no insight to the structures hexagons originated from. In view of the very interesting  properties of the graphene, where the vertices of hexagons are populated by carbon atoms \cite{dres96,har99}, one is motivated to look for the structures from which the original hexagon arose. 

There exist other 2D monolayer materials, some of which are also based on hexagonal lattice (molybdenum disulphide, boron nitride) as well as carbon nanotubes, fullerenes, and heterostructures, that is multi-layered structures build from layers of graphene and other materials \cite{multi1,multi2,multi3}. Is it conceivable that new materials of similar structure will be found in the future. 

In this paper we use group theory to explain how to build a hexagon and hexagonal tiling of a plane by the means of  reflection groups associated with simple Lie groups $SU(3)$, $O(6)$, $O(7)$, $G(2)$, $Sp(6)$. 
Throughout the paper, we use the standard notation for Lie algebras rather than Lie groups, namely, $SU(3)$ is replaced by $A_2$, $G(2)$ by $G_2$, $O(6)$ by $A_3$, $O(7)$ by $B_3$, and $Sp(6)$ by $C_3$.


Distinguished among these possibilities is the pair of cases with $A_2$ and $G_2$ symmetries where construction of hexagonal tiling is planar from the beginning. 
The $A_3$, $C_3$ and $B_3$ algebras are used to produce hexagons which are in 3-dimensional space. Then we list projection matrices that transform 3D hexagons into planar hexagons of $A_2$.

The tool that allows us to consider 3D cases in parallel is the projection matrix from the root space of the algebra $G$ to the root space of its subalgebra $G'$ \cite{computers,NPST}. In all cases, corresponding bases are called $\om$-bases. 

Since we are projecting the orbits of the reflection groups resulting in one or several orbits of the smaller reflection group, the projection matrices listed in the paper are not unique. Indeed, the orbits are invariant with respect to the reflection group of $G$ on one side, and the reflection group of $G'$ on the other. Therefore, the projection matrices may be modified by the transformation of Weyl groups of $G$ into $G'$.


\section{Construction of the graphene sheet}\

Let us recall some facts and set the notation. Each Lie algebra has its own root and weight system and corresponding bases, that is $\a$-basis of (simple) roots and $\om$-basis of (fundamental) weights. $\a$- and $\om$-bases are dual to each other, the correspondence between  is given by the Cartan matrix $C$ and quadratic form matrix $C_q$ \cite{Bourbaki,hum,pinkbook}. 
Both matrices $C$ and $C_q$ are defined as
\begin{align}
(C(G))_{ij}&=(\langle \a_i,\a_j\rangle) =\left( \frac{2(\a_i,\a_j)}{(\a_j,\a_j)} \right), \\
(C_q(G))_{ij}&=(\langle \om_i,\om_j\rangle), \qquad \qquad \qquad i,j=1,\dots , n=rank(G). 
\end{align}
where $\a_i,\a_j$ are simple roots  and $\om_i,\om_j$ are fundamental weights of $G$.
In this paper we use Bourbaki convention for labelling basis elements (simple roots and fundamental weights).
If all roots in the system are of the same length we call them long. For every long root $\a$  we set $\langle \a,\a\rangle =2$.

\subsection{Lie algebra $A_2$ and the graphene}\




There are two ways to construct the graphene sheet from the triangular lattice of Lie algebra $A_2$. One method is based on root lattice $Q$ and the other on  weight lattice $P$. 

Each point of root lattice  $Q$ of $A_2$ is of the form $\Z\a_1 +\Z\a_2$, where $\a_1, \a_2$ are simple roots and $\Z$ are any integer numbers.  Points of the weight lattice $P$ of $A_2$ are of the form $\Z\om_1 +\Z\om_2$, where $\om_1, \om_2$ are fundamental weights.

In the first method, the graphene sheet is formed by the Brillouin zones of the points of root lattice $Q$. The Brillouin zone is also called proximity cell, or in physics literature, {\it Wigner-Seitz cell} or {\it Voronoi domain} \cite{DD, MP92}. 


In other words, the graphene sheet is obtained by assigning to every point of $Q$ its Brillouin zone, that is, the area closer to a given point of $Q$ than to any other point of $Q$. Every Brillouin zone of a point of $Q$ is a regular hexagon. Preserving the hexagons and removing the points of the root lattice from the centers of the hexagons yields the graphene sheet. The Brillouin zones for lattices of all simple roots were described in \cite{MP92}.


\begin{figure}[h] 
\centering\includegraphics[scale=0.1]{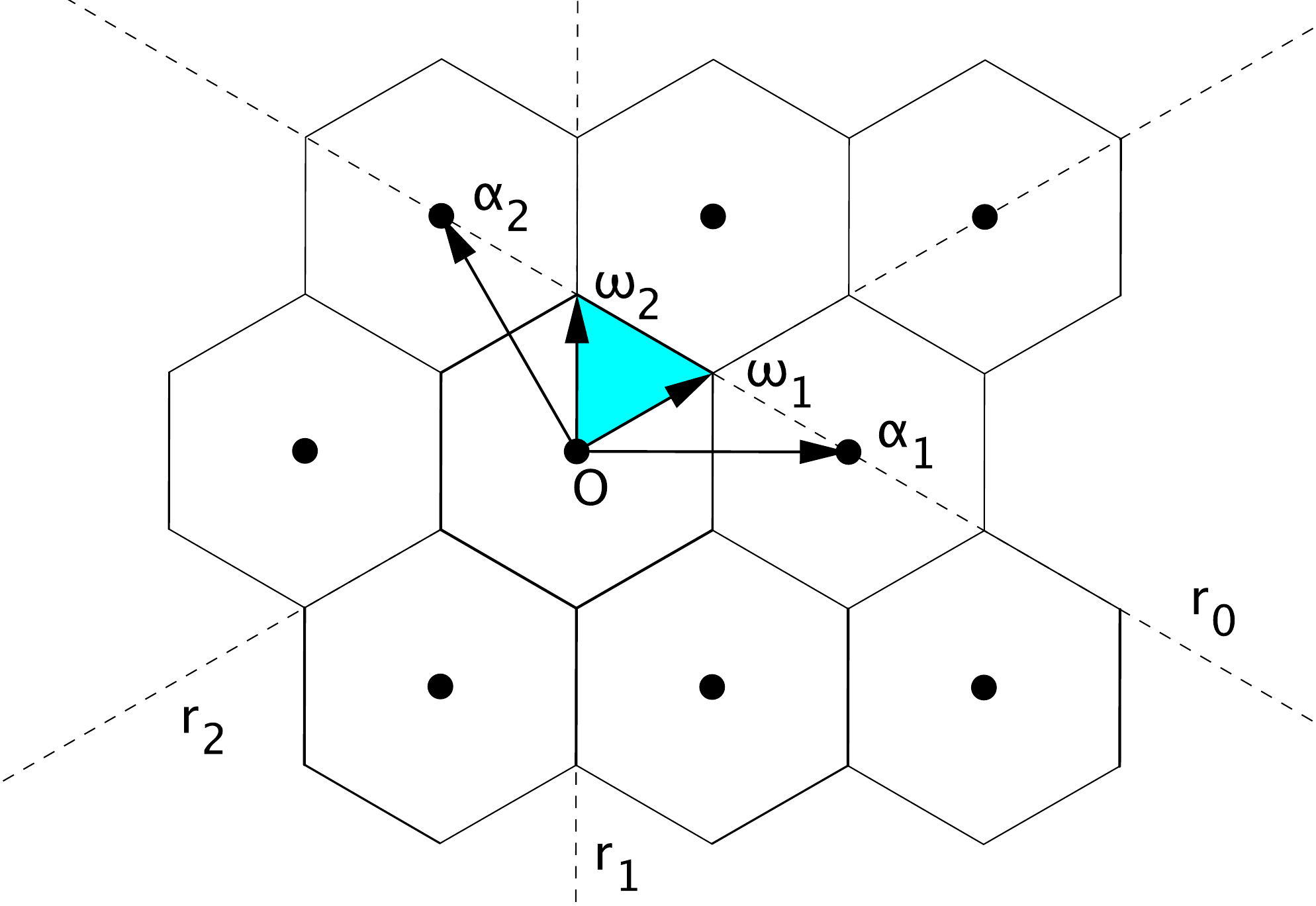}
\caption{Fragment of the root lattice $Q$ lattice of $A_2$ where the Brillouin zones are shown and points of $Q$ are centers of obtained hexagons.}\label{Wigner}
\end{figure}

In the second method we define three congruence classes of points of weight lattice $P$. A  point $a \omega_1 +b \omega_2 \in P$ belongs to one of the three congruence classes $K_i = \{a+2b\equiv i \mod 3\}, i=0,1,2, \  a,b\in
\mathbb{Z}$.

\begin{figure}[h]
\centering\includegraphics[scale=0.15]{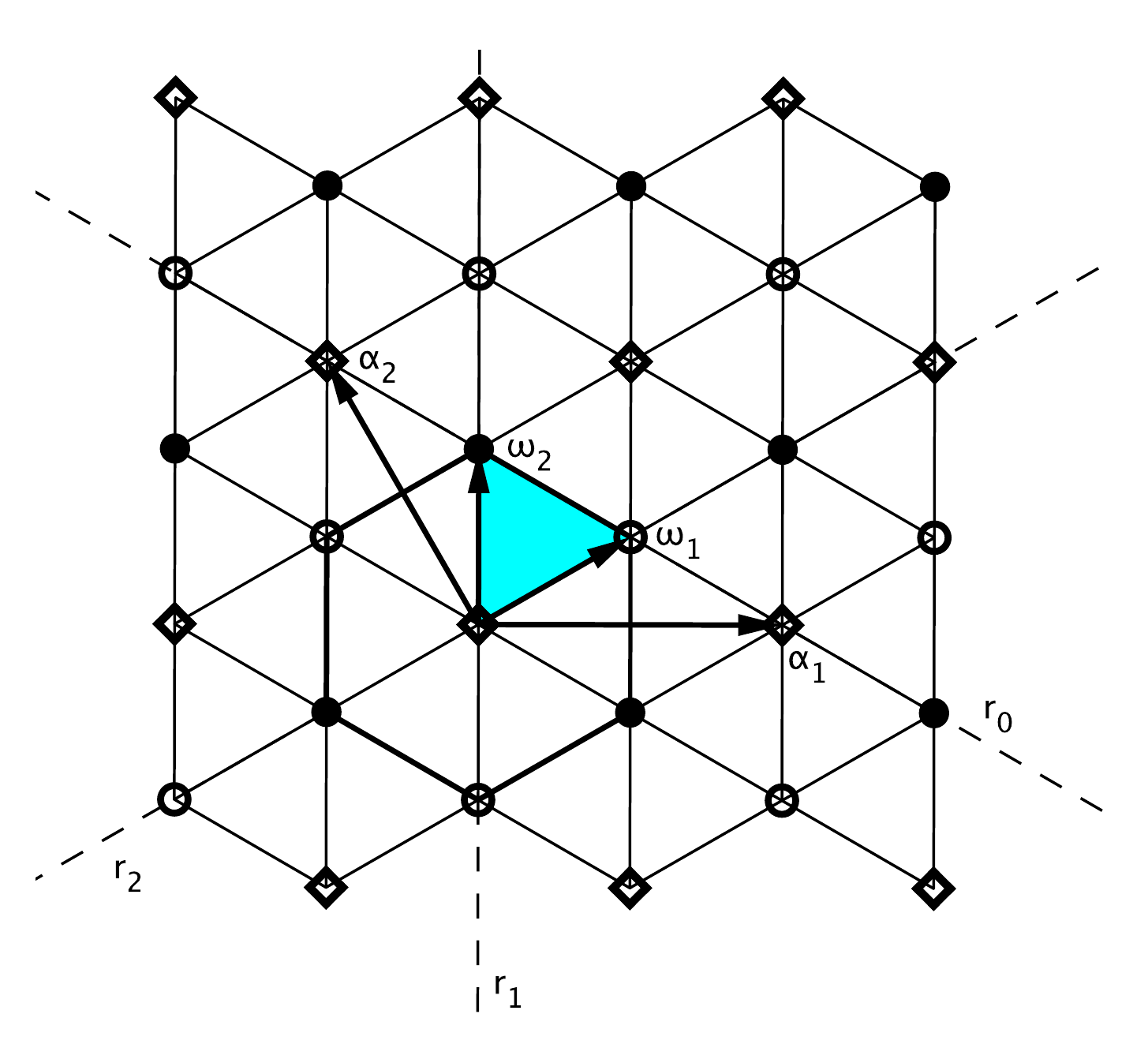}
\caption{Fragment of the $P$ lattice of $A_2$ is shown. Points of congruence class $K_1$ are marked by circles, points of congruence class $K_2$ are marked by black dots, and points of congruence class $K_0$ are marked by rectangles.}\label{classes}
\end{figure}

The triangular lattice $P$ is the union of the points of the three congruence classes $K_i$. Moreover,  points of the  class $K_0$ coincide with the points of root lattice $Q$, points of $K_1$ are formed as $\omega_1+Q$, that means $\om_1$ is added to every point of $Q$, points of $K_2$ are formed as $\omega_2 + Q$. The vertices of the hexagons forming the graphene sheet are obtained by removing from $P$ points of the congruence class $K_0$. 
Each hexagon of the graphene has three vertices in the congruence class $K_1$ and three vertices in the congruence class $K_2$ (see Fig.~\ref{classes}). The origin of the graphene sheet is chosen as the center of one of the hexagons. The center clearly belongs to congruence class $K_0$. 

More generally one can remove any one of the three congruence classes and remaining set of points will yield a graphene sheet.

\subsection{Reflections generating the affine Weyl group of $A_2$}\

As shown in previous section graphene sheet is constructed from lattices of $A_2$. Another possibility is to construct a hexagon from the weight (or root) system of $A_2$ and obtain a graphene sheet by applying the affine Weyl group of $A_2$ to that hexagon.

The Weyl group of $A_2$ is generated by reflections $r_1,r_2$ in the hyperplanes orthogonal to the simple roots $\a_1, \a_2$ of $A_2$ and reflecting mirrors passing through the origin. Explicit reflection formula is given as
\begin{equation}
r_{i}(x)=x-\langle x,\a_1\rangle \a_i=x-\frac{2(x,\a_i)}{(\a_i,\a_i)}\a_i, \quad i=1,2
\end{equation}
where $\a_1,\a_j$ are simple roots, $x$ is any point in the Euclidean plane.

\begin{figure}[h]
\centering\includegraphics[scale=0.15]{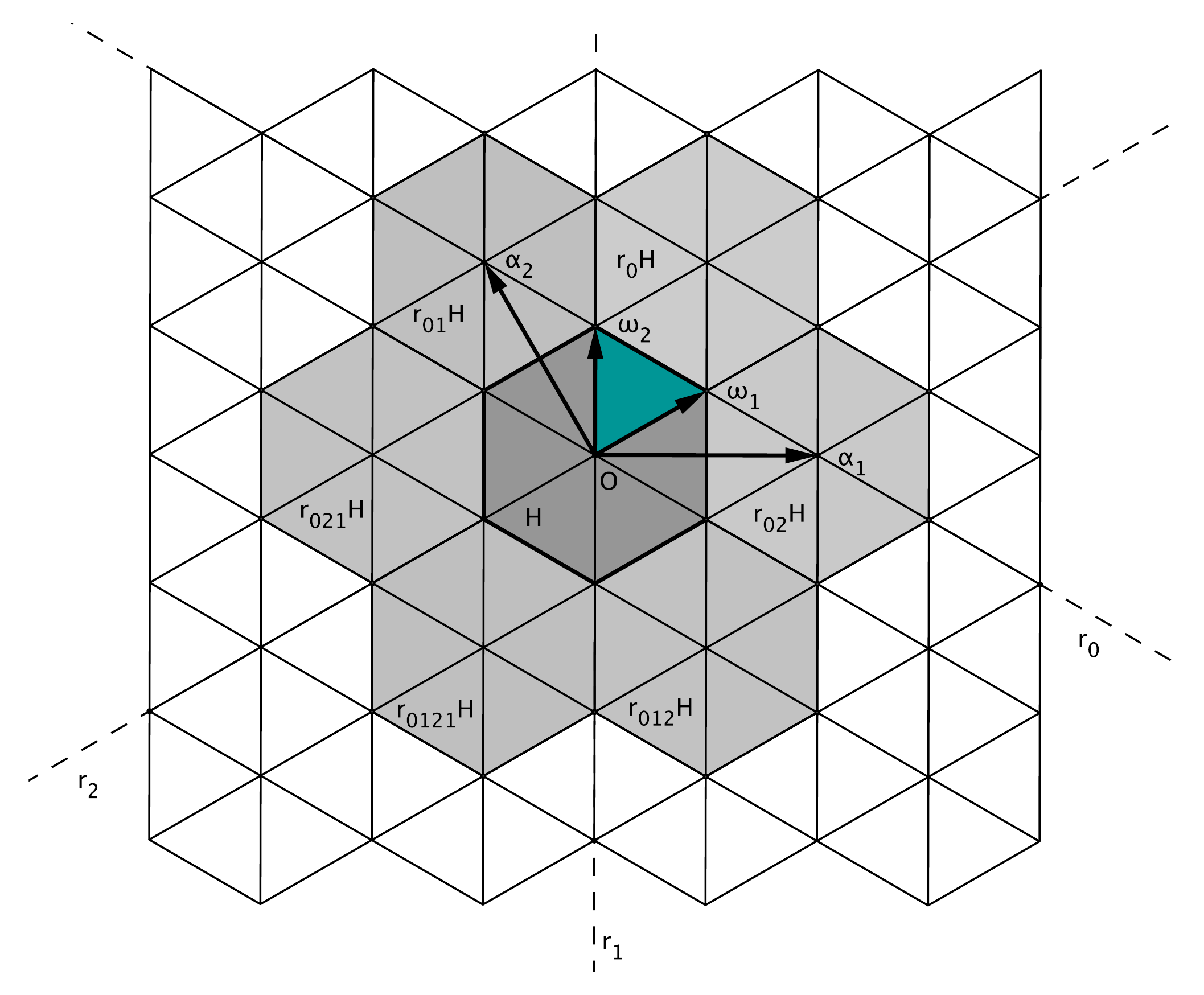}
\caption{Hexagon $H$ and its neighbours, each shown with the reflections used to obtain them. Notation $r_{ij}H$ shows the reflections needed to get the hexagon. First, we apply $r_i$, then $r_j$ to $H$.}\label{hexref}
\end{figure}

We define the affine reflection as 
\begin{equation}
r_{0}(x)=\a_0 + x-\frac{2(x,\a_0)}{(\a_0,\a_0)}\a_0
\end{equation} where $\a_0=\a_1+\a_2$ is the highest root of $A_2$. The affine reflecting mirror does not pass through the origin (see Fig.~\ref{hexref}) and the affine Weyl group is of infinite order.

\subsection{Lie algebra $G_2$ and the graphene}\

Conventional representation of $G_2$ is the one where two simple roots are of different lengths (although the long root could be represented as a linear combination of two short roots). 
The important fact about $G_2$ is that points of its root lattice $Q$ are also the points of its weight lattice $P$. Therefore we cannot introduce the congruence classes of the points of weight lattice $P$ (there is only congruence class 0).

Note that when we consider only long roots of $G_2$ we see that the root system is the same as in the $A_2$ case. Moreover the long simple root of $G_2$ can be presented as a linear combination of its short roots, that means each long root of $G_2$ can be presented in such way, therefore its root lattice $P$ coincides with the lattice of short roots. We can then obtain the hexagonal tiling by considering the $G_2$ lattice of short roots and assigning to each point its Brillouin zone as in the $A_2$ case. 






The second method, based on projection matrices, is more general, and can be applied to higher dimensional Lie algebras. 
The only exception is the chain of subalgebras $B_3\supset G_2\supset A_2$, where $G_2$ is maximal in $B_3$ and $A_2$ is maximal in $G_2$. This is the exceptional inclusion of $G_2$, which has analogs only in much higher ranks of Lie algebras \cite{dynkin}.
The exceptional position of $G_2$ was noted years ago by Racah \cite {Racah}. 

Consider the case of the $G_2$ algebra. Starting from the irreducible 7 dimensional representation of $G_2$, which is the lowest representation of dimension $>1$, the representation of $G_2$ can be reduced to the direct sum of three representations of $A_2$. The reduction process from $G_2$ to $A_2$ consists of projecting the weights of  $G_2$ representation into weights of $A_2$ representations. This is an orthogonal projection provided by $2 \times 2$ matrix found in \cite{computers}, namely
\begin{equation}\label{g2}
Pr(G_2\to A_2)=\left(\begin{matrix}1&1\\1&0
\end{matrix}\right).
\end{equation}
The result of the projection are two triangular orbits of $A_2$, and the origin, which is an orbit in itself, see Example \ref{exm1}. Making the centers of $A_2$ triangles coincide, we obtain a regular hexagon, plus the center point. Removing the center point, we have a hexagon, which, by repeated reflections of the affine Weyl group of $A_2$, gives us the graphene. The projection is done for the $G_2$ weights, which are written in the $\om$-basis of $G_2$, namely 
\begin{equation}\label{g2orb}
\{(0,1), (1,-1), (-1, 2), (1,-2), (-1,1), (0,-1) \}
\end{equation}
and result is in the $\om$-basis of $A_2$.

\begin{example}\label{exm1}
Let us apply the projection matrix \eqref{g2} to the lowest non-trivial $G_2$ orbit \eqref{g2orb} (orbit of the lowest dominant weight).
The projection matrix $Pr(G_2 \to A_2)$ applied to the weight $(0,1)$ of $G_2$ is equal to 
$$\left(\begin{matrix}1&1\\1&0
\end{matrix}\right)
\left(\begin{matrix}0\\1
\end{matrix}\right)= \left(\begin{matrix}1\\0
\end{matrix}\right)$$
weight of $A_2$.
Similarly, we project the remaining weights of the orbit points
\begin{table}[h]
\begin{tabular}{ccccc}
$\left(\begin{smallmatrix}1&1\\1&0
\end{smallmatrix}\right)
 \left(\begin{smallmatrix}1\\-1
\end{smallmatrix}\right)= \left(\begin{smallmatrix}0\\1
\end{smallmatrix}\right),$ & \qquad & 
$\left(\begin{smallmatrix}1&1\\1&0
\end{smallmatrix}\right)
 \left(\begin{smallmatrix}-1\\2
\end{smallmatrix}\right)= \left(\begin{smallmatrix}1\\-1
\end{smallmatrix}\right),$& \qquad & 
$\left(\begin{smallmatrix}1&1\\1&0
\end{smallmatrix}\right)
\left(\begin{smallmatrix}1\\-2
\end{smallmatrix}\right)= \left(\begin{smallmatrix}-1\\1
\end{smallmatrix}\right),$ \\
$\left(\begin{smallmatrix}1&1\\1&0
\end{smallmatrix}\right)
 \left(\begin{smallmatrix}-1\\1
\end{smallmatrix}\right)= \left(\begin{smallmatrix}0\\-1
\end{smallmatrix}\right),$ & \qquad & 
$\left(\begin{smallmatrix}1&1\\1&0
\end{smallmatrix}\right)
 \left(\begin{smallmatrix}0\\-1
\end{smallmatrix}\right)= \left(\begin{smallmatrix}-1\\0
\end{smallmatrix}\right).$
\end{tabular}
\end{table}

The result of the projection are the six vertices of the $A_2$ hexagon. Note, however, that this hexagon is formed by two triangular orbits centered at the origin, namely $\{(1,0)$, $(-1,1)$, $(0,-1)\}$ and $\{(0,1)$, $(1,-1)$, $(-1,0)\}$.

\end{example}

\subsection{3-dimensional hexagons and the graphene}\

The rapidly increasing use of the graphene sheets in large variety of situations including other atoms than carbon leads one to the idea that such sheets could turn out to be not strictly planar, or becoming planar after some additional projection to a plane where they become graphene.
In this section we brought up somewhat speculative topic of forming sheets of hexagons, but sheets that may not be in a plane.  For that we recall the hexagonal orbits of the Lie algebras $A_3$, $B_3$, $C_3$. 

\begin{figure}[h]
\includegraphics[scale=0.35]{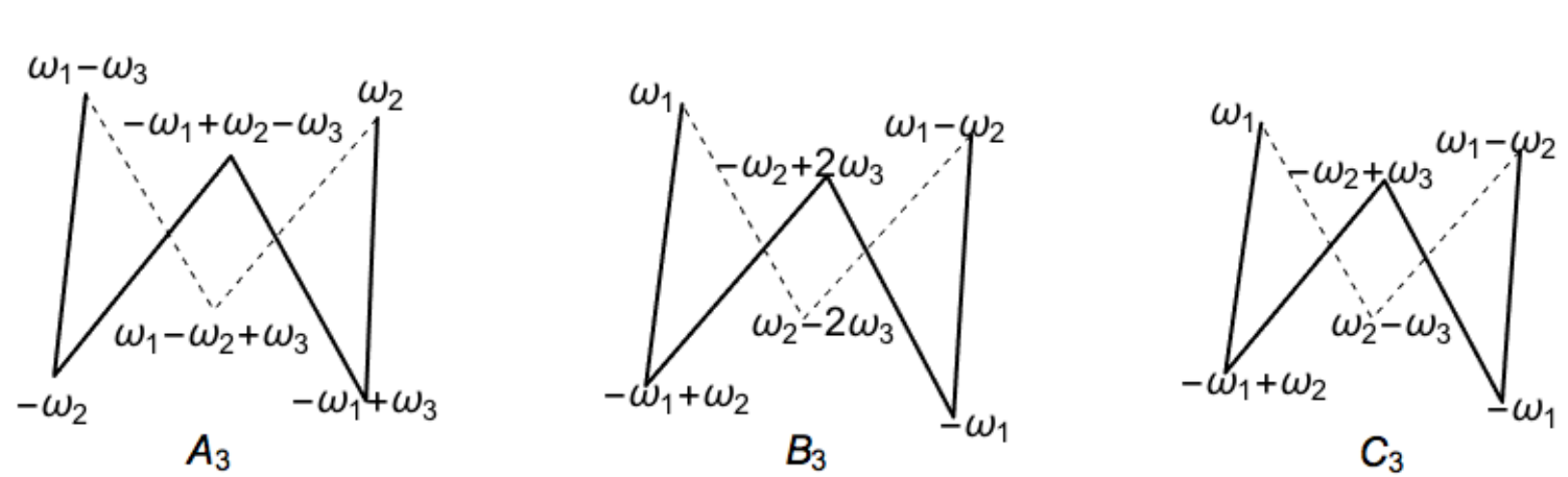}
\caption{Nonplanar hexagons with vertices in the coordinates of $\om$-bases of  $A_3, B_3$ and $C_3$.}\label{hexa3D}
\end{figure}

The subsequent three cases described in this paper are formed as follows. We consider rank 3 Lie algebras where $A_2$ is a maximal subalgebra, that is $A_3$, $B_3$ and $C_3$. In those cases the lowest irreducible representation contains a non-planar hexagonal orbit of weights (Fig.~\ref{hexa3D}), which can be subsequently projected by known projection matrices into the hexagons of $A_2$. Each hexagon is then reflected in its sides by the affine Weyl group of $A_2$, so that it is spread all over the plane, forming the graphene.

Curiously, the 3 hexagons of $A_3$, $B_3$ and $C_3$ in 3D are identical as geometric structures. Their vertices are labelled in relative $\om$-bases, as shown in Fig.~\ref{hexa3D}.
 
Consider the weight system of the irreducible representation of $A_3$ with the highest weight $(0,1,0)$ in $A_3$ $\om$-basis. The three reflections in the hyperplane orthogonal to the simple roots of $A_3$, and intersecting in the origin applied to the highest weight, yield the hexagon with vertices 
 \begin{equation}
\{ (0,1,0), (1,-1,1), (1,0,-1), (0,-1,0), (-1,1,-1), (-1,0,1) \} 
 \end{equation}
 
The projection matrix 
\begin{equation}\label{a3}
 Pr(A_3 \to A_2)=\left(\begin{matrix}
1 & 1 & 0 \\
 0 &0 &1
\end{matrix}\right)
\end{equation}
transforms the $A_3$ hexagon (Fig.~\ref{hexa3D}) into the planar $A_2$ hexagon in its $\om$-basis, namely 
\begin{equation}
\{(1,0), (0,1),(1,-1),(-1,0),(0,-1),(-1,1)\}.
\end{equation}
 
The Lie algebra $B_3$ has the lowest irreducible representation of dimension 7. The weight system of the representation consists of two orbits of the $B_3$ Weyl group. One orbit of six points forming a hexagon, and one orbit which is the origin. There are two possible chains of maximal subalgebras, which start with $B_3$ and end in $A_2$. The first is $B_3 \supset G_2 \supset A_2$, and the second is $B_3 \supset A_3 \supset A_2$. The non-planar $B_3$ hexagon, see Fig.~\ref{hexa3D}, has the following vertices
\begin{equation}
\{(1,0,0), (-1,1,0), (0,-1,2),(-1,0,0),(1,-1,0),(0,1,-2)\}.
\end{equation}

It can be first projected into the $G_2$ hexagon or into the $A_3$ hexagon using the appropriate matrix
\begin{equation}
Pr(B_3 \to G_2)=\left(\begin{matrix}
0& 1 & 0 \\
1 & 0 &1
\end{matrix}\right) \quad \quad
Pr(B_3 \to A_3)=\left(\begin{matrix}
0&1&1\\
1&0&0\\
0&1&0
\end{matrix}\right)
\end{equation}
and then into the $A_2$ hexagon using either the \eqref{g2} or \eqref{a3} matrix. Combining this two-step projection, we get
\begin{equation}
Pr(B_3 \to A_3 \to A_2)=Pr(B_3 \to G_2 \to A_2)=\left(\begin{matrix}
1&1&1\\
0&1&0
\end{matrix}\right).
\end{equation}
In both cases, $B_3$ to $A_2$ projection is done using the same matrix. It remains to apply the affine Weyl group of $A_2$ to obtain the hexagon in order to produce the graphene.

\begin{example}

Let us demonstrate the coincidence of the projection matrices of $B_3$ to $A_2$ when proceeding with different chains of subalgebras. Consider $B_3 \supset G_2 \supset A_2$. Then
$$Pr(G_2 \to A_2)Pr(B_3 \to G_2)=
\left(\begin{matrix}
1& 1  \\
1 & 0
\end{matrix}\right)
\left(\begin{matrix}
0& 1 & 0 \\
1 & 0 &1
\end{matrix}\right)=\left(\begin{matrix}
1&1&1\\
0&1&0
\end{matrix}\right).$$

Consider $B_3 \supset A_3 \supset A_2$. We get
$$Pr(A_3 \to A_2)Pr(B_3 \to A_3)=
\left(\begin{matrix}
1 & 1 & 0 \\
 0 &0 &1
\end{matrix}\right)
\left(\begin{matrix}
0&1&1\\
1&0&0\\
0&1&0
\end{matrix}\right)=\left(\begin{matrix}
1&1&1\\
0&1&0
\end{matrix}\right).
$$
\end{example}

Consider the Lie algebra $C_3$. Its lowest irreducible representation is of dimension 6. The corresponding weight system consists of six weights
\begin{equation}
\{(1,0,0), (-1,1,0), (0,-1,1),(-1,0,0),(1,-1,0),(0,1,-1)\}
\end{equation}
in $C_3$ $\om$-basis. The orbit forms a non-planar hexagon, see Fig.~\ref{hexa3D}.
The projection matrix
\begin{equation}
Pr(C_3\to A_2)=\left(\begin{matrix}
1&1&2\\
0&1&0\\
\end{matrix}\right)
\end{equation}
transforms the hexagon of $C_3$ into the same $A_2$ hexagon as in the previous cases. 

\section{Colouring the graphene}\ 

It is conceivable that in some near future applications some hierarchy of the hexagons of the graphene may need to be introduced. The simplest example we can imagine at this point is the multilayered graphene where the orientation of hexagons on each layer is governed by some additional rule. 

In this section, we introduce such hierarchy of the hexagons in a graphene sheet. We call it \emph{colouring}, for lack of a better name. In addition, we also consider transformations between different colourings, which we refer to as phase transitions. 

There is an algebraic rule for assigning an integer (colour) to each hexagon. Practically, we proceed as follows. The center of each hexagon is a point of congruence class $K_0$ of the weight lattice $P$ of $A_2$. It also happens to be a point of the root lattice $Q$ of $A_2$. We therefore assign a colour to each point of $Q$, and recognize that this is the colour of the whole hexagon containing the point. The number of colours can be any positive integer, but in this paper we focus on 2 and 3 colours in details.

A point of $Q$ is of the form $a\alpha_1 +b\alpha_2 $, where $a,b \in \mathbb{Z}$. The colour $(k_1,k_2) \mod m$ of a point is defined as
\begin{equation}\label{colour}
\left(\begin{matrix}
a & b
\end{matrix}\right)
\left(\begin{matrix}
k_1\\
k_2
\end{matrix}\right) = k_1 a +k_2 b \mod m ,
\end{equation}
where the positive integer $m$ is the number of colours, and the pair $(k_1, k_2) \mod m$ specifies the colouring type. The trivial colouring is given by $(k_1,k_2)=(0,0)$.

Suppose we have a coloured graphene of the type $(k_1,k_2) \mod m$. We can represent the colouring as a column vector $(k_1, k_2,1)^T$. The transition to another colouring characterized by integers $(l_1,l_2)\mod m$ is given by the matrix
\be T=\left( \begin{matrix}
1&0&l_1\\0&1&l_2\\
0&0&1
\end{matrix}\right). \ee Namely, 
\be\left( \begin{matrix}
1&0&l_1\\0&1&l_2\\
0&0&1
\end{matrix}\right) \left(\begin{matrix}
k_1\\
k_2\\1
\end{matrix} \right)=
\left(
\begin{matrix}
k_1+l_1\\ 
k_2+l_2\\
1
\end{matrix}\right).\ee
Composition of several transitions is possible, through a  product of matrices i.e.,
\begin{equation}\label{product}
\left( \begin{matrix}
1&0&l_1\\0&1&l_2\\
0&0&1
\end{matrix}\right)\left( \begin{matrix}
1&0&c_1\\0&1&c_2\\
0&0&1
\end{matrix}\right)=
\left( \begin{array}{ccc}
1 & 0 & l_1+c_1\\
0&1&l_2+c_2\\
0&0&1
\end{array}\right).
\end{equation}
Clearly, for every transition matrix $T$, there exists its inverse
$$\left( \begin{matrix}
1&0&l_1\\0&1&l_2\\
0&0&1
\end{matrix}\right)\left( \begin{matrix}
1&0&-l_1\\0&1&-l_2\\
0&0&1
\end{matrix}\right)=
\left( \begin{array}{ccc}
1 & 0 & 0\\
0&1&0\\
0&0&1
\end{array}\right).$$
Therefore, the phase transition matrices $T$ form a finite Abelian group whose order is controlled by the number of colours $m$.

\begin{example}
2-colourings of the graphene.

There are 3 non-equivalent ways to assign 2 colours to all the points of $Q$. In other words, there are 3 different non-trivial 2-colourings of the graphene sheet. Namely,
\begin{align*}
 (k_1,k_2) = (1,0), \qquad   (k_1,k_2)=(0,1), \qquad  (k_1,k_2)=(1,1).
\end{align*}

\begin{figure}[h]
\centering\includegraphics[scale=0.3]{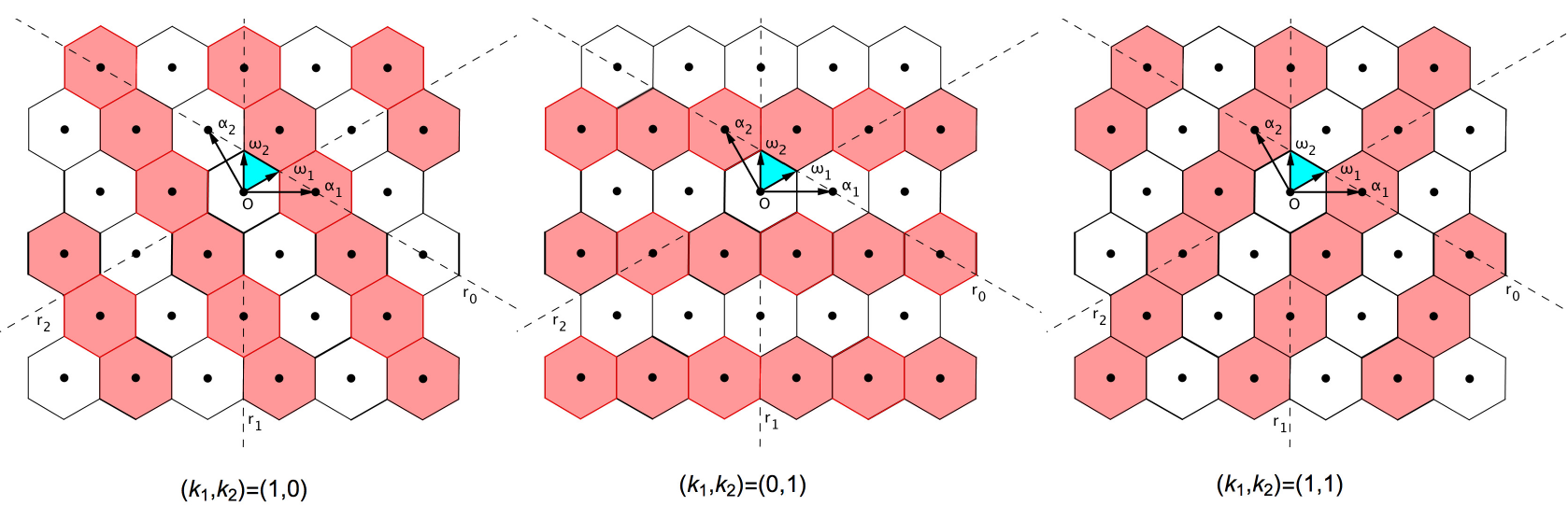}

\caption{Examples of fragments of the graphene with 2-colouring of the hexagons.}\label{2col}
\end{figure}

\noindent Any other choice of integers $(k_1,k_2)$ turns out to be equivalent to one of the given three, because we are considering the colouring $\mod 2$. Examples of the three types of such colouring are shown in Fig.~\ref{2col}.
 \end{example}
 
\begin{figure}[h]
\centering\includegraphics[scale=0.3]{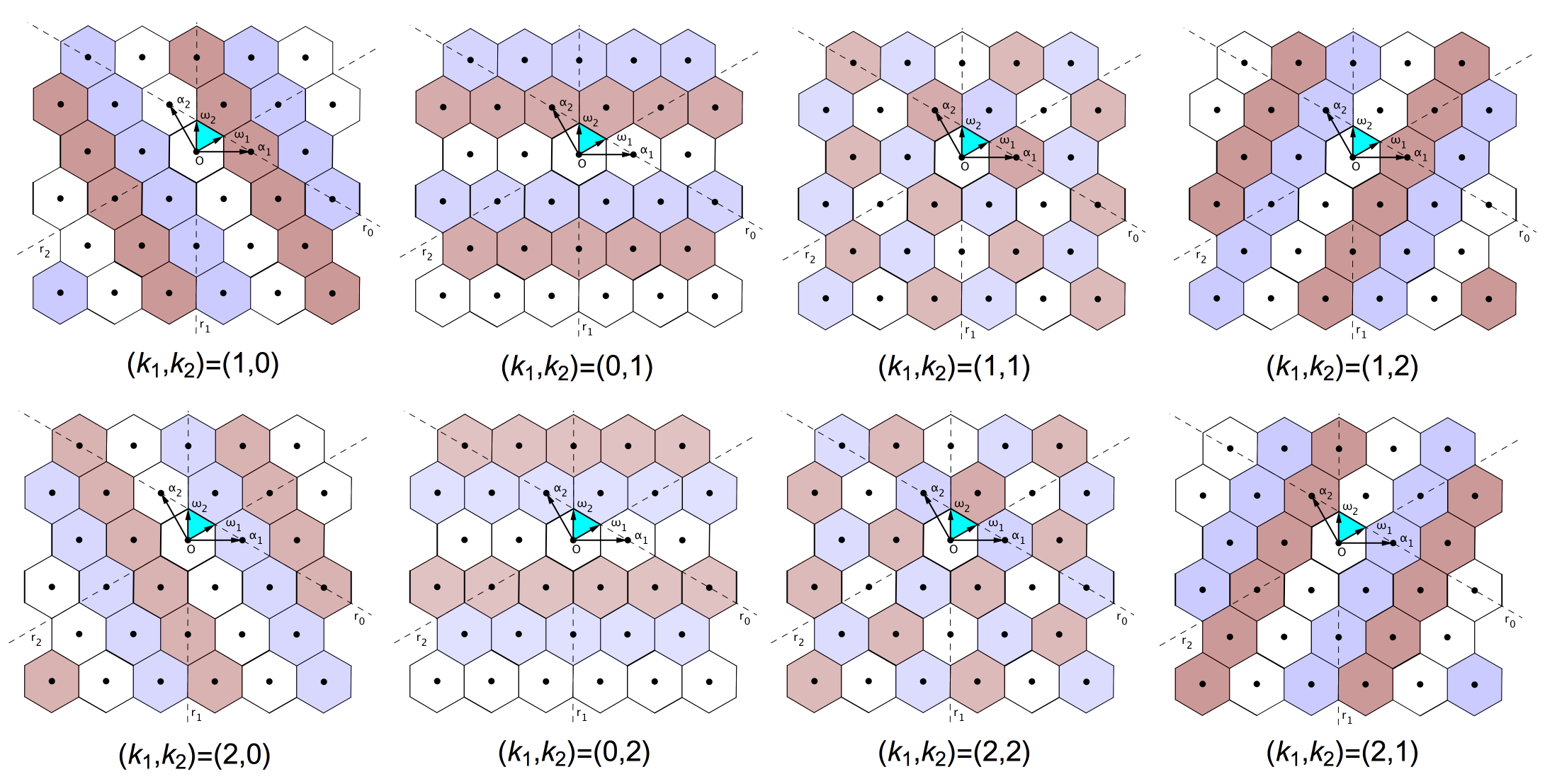}

\caption{Examples of fragments of the graphene with 3-colouring of  hexagons.}\label{3col}
\end{figure}

\begin{example}
3-colourings of the graphene.

There are 8 non-trivial choices of integers $(k_1, k_2 )\mod 3$ that specify non-equivalent 3-colourings of the graphene's hexagons. These are the following:
\begin{align*}
& (k_1,k_2) = (1,0), \qquad  (k_1,k_2) = (0,1), \qquad  (k_1,k_2)=(1,1), \\
&(k_1,k_2) = (2,0), \qquad  (k_1,k_2)=(0,2), \qquad (k_1,k_2) = (2,2), \\
& (k_1,k_2)=(1,2), \qquad  (k_1,k_2)=(2,1).
\end{align*}
Examples of such colourings are shown in Fig~\ref{3col}.

\end{example}

\noindent Analogously, colouring can be done for an arbitrary $m<\infty $.

\section{Refinement of the graphene}\


A straightforward way to refine the graphene is to cut each of its hexagons into smaller hexagons. In this case, admissible sizes of these hexagons must be set. In the paper, we present two more systematic ways to refine the graphene.

Consider the graphene and the vertices of each hexagon. Brillouin zones of the vertices of hexagons form a triangular lattice $P'$ in $\mathbb{R}^2$. These triangles are smaller than those in the original weight lattice $P$ of $A_2$. Three congruence classes of points of $P'$ can be defined. The graphene of $P'$ is obtained by deleting the congruence class zero from the lattice $P'$. This process can be repeated as many times as desired, creating a graphene of smaller and smaller hexagons at each step.

\begin{figure}[h]
\centering\includegraphics[scale=0.1]{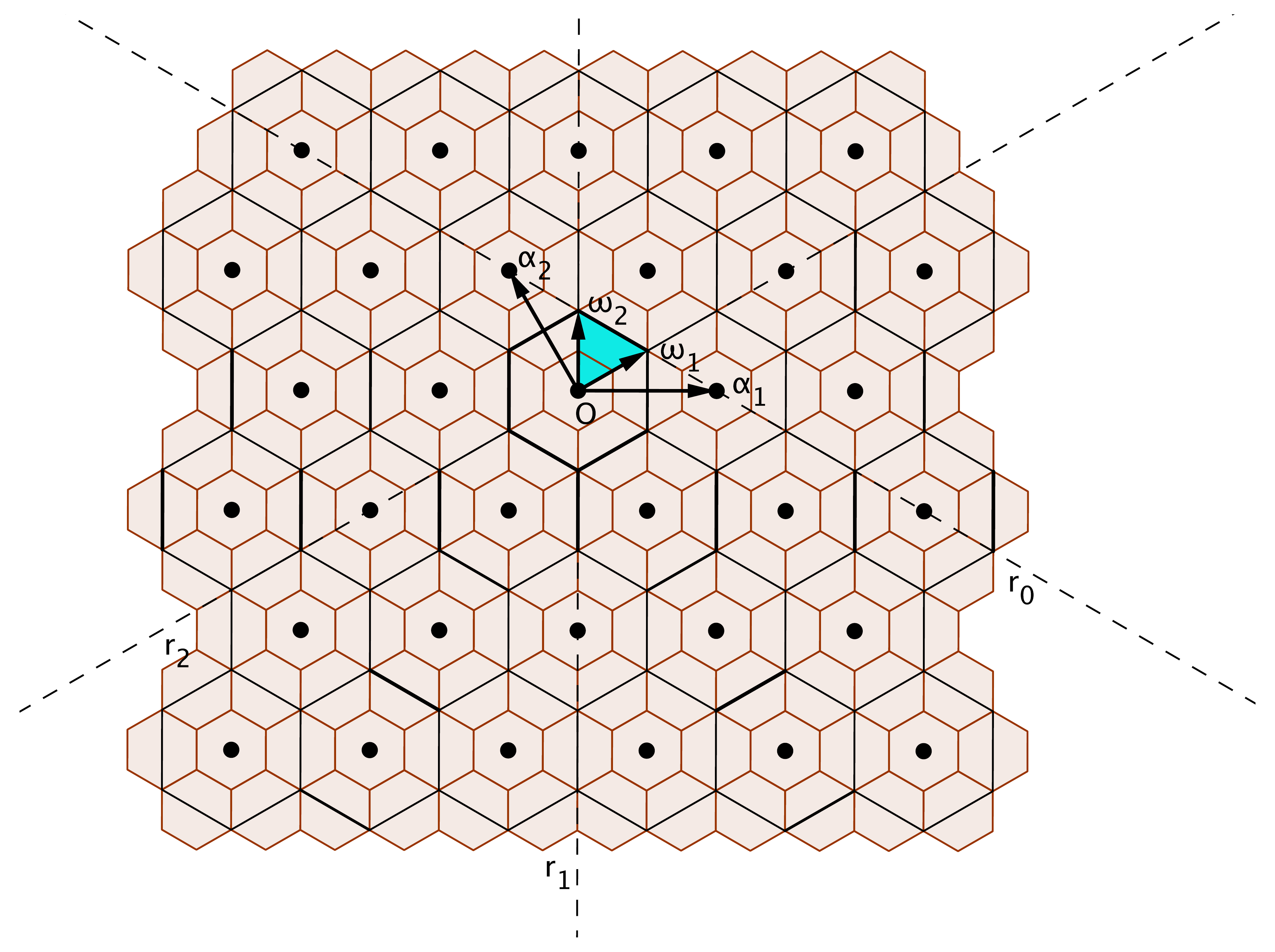}
\caption{Fragment of the refinement of a graphene sheet.}\label{ref}
\end{figure}

The second method is more powerful, as well as more general. In this paper, it is only used for the Lie algebra $A_2$. The refinement of the graphene starts with the refinement of the lattice $P$ of $A_2$. The basic tile $F$ in the lattice $P$ is the coloured triangle in Fig.~\ref{classes}. Its vertices are $0,\om_1,\om_2$. Fix a positive integer $M$. We obtain $|F_M (A_2)| = \left( \begin{matrix}
M+2\\
2
\end{matrix}\right)$
points in the basic tile and on its boundary, by providing the 3 non-negative integers $[s_0,s_1,s_2]$ such that 
\begin{equation}\label{sum}
s_0+s_1+s_2=M.
\end{equation} 


The coordinates of the points in $F_M$ are $\frac{s_1}{M}\om_1 +\frac{s_2}{M}\om_2$. When one considers all possible choices for $s_i$ integers such that \eqref{sum} holds, one obtains the refinement of the basic tile, where the points are vertices of the triangles. Reflections from the affine Weyl group of $A_2$ spread the refinement onto the whole plane. Taking the refined lattice, we can start to build the graphene as in section 2.1, and obtain a refined graphene whose density is fixed by $M$.


\section{Concluding remarks}\


Thus far, we have considered the possibility of constructing the $A_2$ graphene by means of the inclusion of $A_2$ as a maximal subalgebra of some higher rank algebras. A few additional possibilities exist. Rather than making $A_2$ a maximal subalgebra of a simple Lie algebra, we can require that $A_2$ is a maximal subjoint algebra \cite{slansky}.

The method for constructing the graphene sheet can be generalized to more dimensions in a straightforward way. Indeed, generalization of the graphene sheets built using Brillouin zones is possible for any dimension and any symmetry group. The classification of Brillouin zones is found in \cite{MP92}. 

Generalization of the construction of graphenes by means of congruence classes is possible only when the symmetry is of type $A_n$, $n<\infty $. Higher dimensional graphene-like structures are not formed by hexagons in general.



Colouring of graphenes can be generalized to higher dimensional lattices of simple Lie algebras. Moreover, the phase transition can be generalized to more dimensions. 

It would be interesting to describe finite Abelian groups formed by the phase transition matrices for more general setups. 

Possibilities of colouring the graphene aperiodically remain unexplored. The simplest way to achieve this would be to map $Q$ to the 2-dimensional quasicrystal.

Many properties of the graphene, including colouring, can be extended to nanotubes, which are strips of the graphene of appropriate width, rolled into a cylinder \cite{cal,kit}. Different nanotubes are obtained by different orientations of the strips \cite{ 
BPS13,BPS14,BBPS14,BBPS15}.

\subsection*{Acknowledgements}

Authors are grateful to the Natural Sciences and Engineering Research Council of Canada, RGPIN-2016-04199.



\address{
$^1$ Centre de recherches math\'ematiques, Universit\'e de Montr\'eal, C.~P.~6128, succ. Centre-ville, Montr\'eal, H3C\,3J7, Qu\'ebec, Canada\\
$^2$ D\'epartement de math\'ematiques et de statistique, Universit\'e de Montr\'eal, C.~P.~6128, Centre-ville, Montr\'eal, H3C\,3J7, Qu\'ebec, Canada\\
$^3$ Institute of Mathematics, University of Bialystok, 1M Ciolkowskiego, PL-15-245, Bialystok, Poland\\
\email{grabowie@dms.umontreal.ca, patera@crm.umontreal.ca, \\m.szajewska@math.uwb.edu.pl}
}

\end{document}